\documentclass[journal]{IEEEtran}

\ifCLASSINFOpdf
\else
   \usepackage[dvips]{graphicx}
\fi
\usepackage{url}
\usepackage{textcomp}
\usepackage{stfloats}
\usepackage{hyperref} 
\usepackage{subfigure}
\usepackage{verbatim}
\usepackage{makecell}
\usepackage{multirow}
\usepackage{graphicx}
\usepackage{cite}
\usepackage{amssymb}
\usepackage{color}
\usepackage{xcolor}
\usepackage{amsmath,amsfonts}
\usepackage{algorithmic}
\usepackage{algorithm}
\usepackage{array}
\usepackage{booktabs}
\begin{document}

\title{LM-VC: Zero-shot Voice Conversion via Speech Generation based on Language Models}

\author{Zhichao~Wang,
    Yuanzhe~Chen,
    Lei~Xie,~\IEEEmembership{Senior Member,~IEEE,}
    Qiao Tian,
    Yuping Wang
\thanks{Corresponding author: Lei Xie, Qiao Tian}
\thanks{Zhichao Wang and Lei Xie are with the ASLP Lab, School of Computer Science, Northwestern Polytechnical University, Xi’an 710129, China (email: zcwang\_aslp@mail.nwpu.edu.cn; lxie@nwpu.edu.cn)}
\thanks{Yuanzhe Chen, Qiao Tian, and Yuping Wang are with the ByteDance SAMI Group, Shanghai 200233, China (email: chenyuanzhe@bytedance.com; tianqiao.wave@bytedance.com, wangyuping@bytedance.com)}
    }

\markboth{Journal of \LaTeX\ Class Files, Vol. 14, No. 8, August 2015}
{Shell \MakeLowercase{\textit{et al.}}: Bare Demo of IEEEtran.cls for IEEE Journals}
\maketitle

\begin{abstract}

Language model (LM) based audio generation frameworks, e.g., AudioLM, have recently achieved new state-of-the-art performance in zero-shot audio generation. In this paper, we explore the feasibility of LMs for \textit{zero-shot voice conversion}. An intuitive approach is to follow AudioLM -- Tokenizing speech into semantic and acoustic tokens respectively by HuBERT and SoundStream, and converting source semantic tokens to target acoustic tokens conditioned on acoustic tokens of the target speaker. However, such an approach encounters several issues: 1) the linguistic content contained in semantic tokens may get dispersed during multi-layer modeling while the lengthy speech input in the voice conversion task makes contextual learning even harder; 2) the semantic tokens still contain speaker-related information, which may be leaked to the target speech, lowering the target speaker similarity; 3) the generation diversity in the sampling of the LM can lead to unexpected outcomes during inference, leading to unnatural pronunciation and speech quality degradation. To mitigate these problems, we propose \textit{LM-VC}, a two-stage language modeling approach that generates coarse acoustic tokens for recovering the source linguistic content and target speaker's timbre, and then reconstructs the fine for acoustic details as converted speech. Specifically, to enhance content preservation and facilitates better disentanglement, a masked prefix LM with a mask prediction strategy is used for coarse acoustic modeling. This model is encouraged to recover the masked content from the surrounding context and generate target speech based on the target speaker's utterance and corrupted semantic tokens. Besides, to further alleviate the sampling error in the generation, an external LM, which employs window attention to capture the local acoustic relations, is introduced to participate in the coarse acoustic modeling through shallow fusion. Finally, a prefix LM reconstructs fine acoustic tokens from the coarse and results in the converted speech. Experiments demonstrate that LM-VC outperforms competitive systems in speech naturalness and speaker similarity.

\end{abstract}
\vspace{-3pt}

\IEEEpeerreviewmaketitle

\vspace{-13pt}
\section{Introduction}

\IEEEPARstart{V}{oice} conversion (VC) aims to convert speech from a source speaker to that of a target speaker without changing the linguistic content. VC's main rationale is to decompose source speech into separated components, including speaker timbre, linguistic content, and speaking style. Then the linguistic content and speaking style are combined with the target speaker's timbre to generate the converted speech. Training a typical VC system desires at least a sizable amount of the target speaker's speech. In contrast, \textit{zero-shot} VC or \textit{any-to-any} VC focuses on converting any source speech to that of any desired speaker with only one utterance available from the speaker, which is more practical for real-world applications. But since only one target speaker utterance is available, decoupling speech components and meanwhile maintaining target speaker timbre becomes more challenging.



One intuitive approach is to leverage a speaker verification (SV) model to extract the speaker representation~\cite{autovcqian2019autovc,speechsplit,mediumvc}, while automatic speech recognition (ASR)~\cite{PPGSun2016PhoneticPF} or self-supervised learning (SSL) model~\cite{nansy,FragmentvcAVLin2021,retriever,speaking} are employed to extract the linguistic content. Some studies~\cite{speechsplit,nansy} also use signal perturbation techniques to alter speech utterances to make it speaker irrelevant before content extraction. Instead of attempting speech disentanglement prior to training the VC model, many studies rely on a specifically designed disentanglement approach to reduce the correlation among different speech components, including designs on complicated neural structures~\cite{INchou2019oneshot,wang2023multilevel}, loss functions~\cite{VQMIVC,MAP}, and training strategies~\cite{avqvc,contrastive}. 
However, the current zero-shot VC approaches still generalize poorly to unseen speakers with low speaker similarity, mainly due to the inevitable trade-off during the speech disentanglement process and the model's limited capacity on leveraging large-scale speech data.

Recently, language models (LM)~\cite{audiolm,musiclm,valle,spearTTS} trained on large-scale datasets have achieved impressive performance in zero-shot audio generation. A popular paradigm is first to tokenize audio into \textit{semantic} and \textit{acoustic} tokens respectively by a self-supervised learning (SSL) model and a neural codec, where the SSL model extracts the linguistic content from audio while the audio codec reconstructs high-quality audio at very low bitrate, and then the discrete tokens enable the audio generation task to benefit from the powerful large language models. As a typical approach, AudioLM~\cite{audiolm} leverages 
semantic and acoustic tokens as audio representations and introduces a three-stage language modeling process for audio generation. Specifically, using semantic tokens of a short utterance as a \textit{prompt}, AudioLM generates the continuation of semantic tokens, which is then used as a conditioning signal for predicting coarse acoustic tokens and further restoring the fine acoustic details. Variants of AudioLM have also shown remarkable performance for zero-shot music generation and text-to-speech (TTS)~\cite{musiclm,valle,spearTTS}.

In this letter, we explore the feasibility of language models in zero-shot VC. 
An intuitive way is to follow AudioLM -- applying coarse and fine acoustic modeling to form a variant. However, such a straightforward LM approach encounters several issues in voice conversion:
1) the linguistic content contained in semantic tokens may get dispersed as the network deepens during multi-layer language modeling while the lengthy speech input makes contextual learning even harder; 2) the semantic tokens extracted by HuBERT~\cite{hsu2021hubert} still contain speaker-related information, which may be propagated to the converted speech and lead to low speaker similarity; 3) the inherent generation diversity in the sampling of LM inevitably leads to unnatural pronunciation and even speech quality degradation. To address these issues, we propose a language model-based VC approach (LM-VC) -- a two-stage framework that first generates coarse acoustic tokens for recovering content and speaker timbre and then reconstructs the fine acoustic details as converted speech. Specifically, to maintain linguistic content and facilitate better speech disentanglement, we use a masked prefixed language model (MPLM) with a mask prediction strategy for coarse acoustic modeling. This model is encouraged to recover masked semantic tokens based on the context and predict target speech given the target speaker's utterance and the corrupted semantic tokens, thereby implicitly creating an information bottleneck on the source speech to reduce the source speaker information. To further alleviate the sampling error in the generation process, we integrate an external language model (ELM) that employs window attention~\cite{liu2021swin} to better capture the local context among acoustic tokens. The ELM collaborates with the MPLM through \textit{shallow fusion}~\cite{shallow1} to generate target speech. Finally, the fine acoustic tokens are reconstructed from the coarse ones in a non-autoregressive manner using a prefix LM~\cite{valle}. Experiments and ablations on large-scale speech data show that LM-VC is superior to YourTTS~\cite{yourtts} and an AudioLM~\cite{audiolm} baseline in both speaker similarity and speech naturalness.

\begin{figure}[ht]
\centering
\vspace{-10pt}
\begin{minipage}{0.65\linewidth}
    \subfigure{
      \includegraphics[width=1\columnwidth]{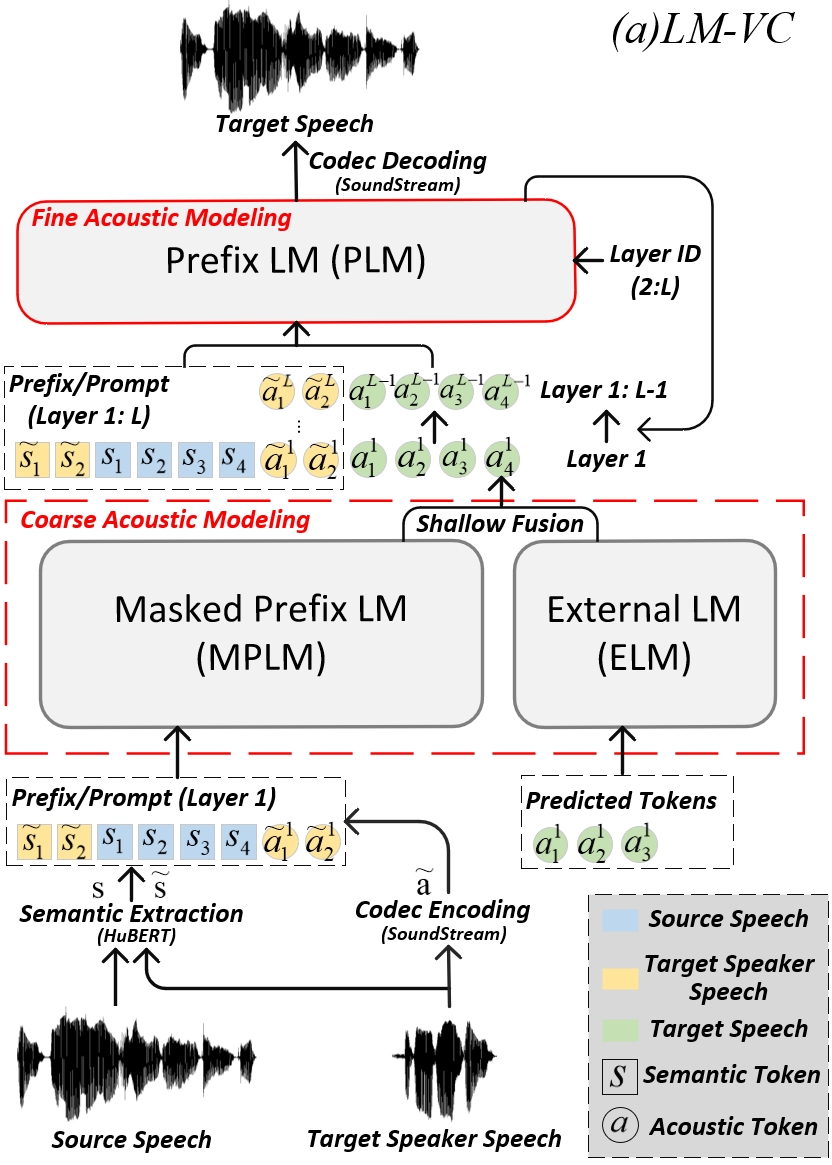}}
\end{minipage}
\textcolor[RGB]{180,180,180}{\rule{0.8\linewidth}{0.7pt}}
\begin{minipage}{0.8\linewidth}
    \vspace{3pt}
    \subfigure{
      \includegraphics[width=1\columnwidth]{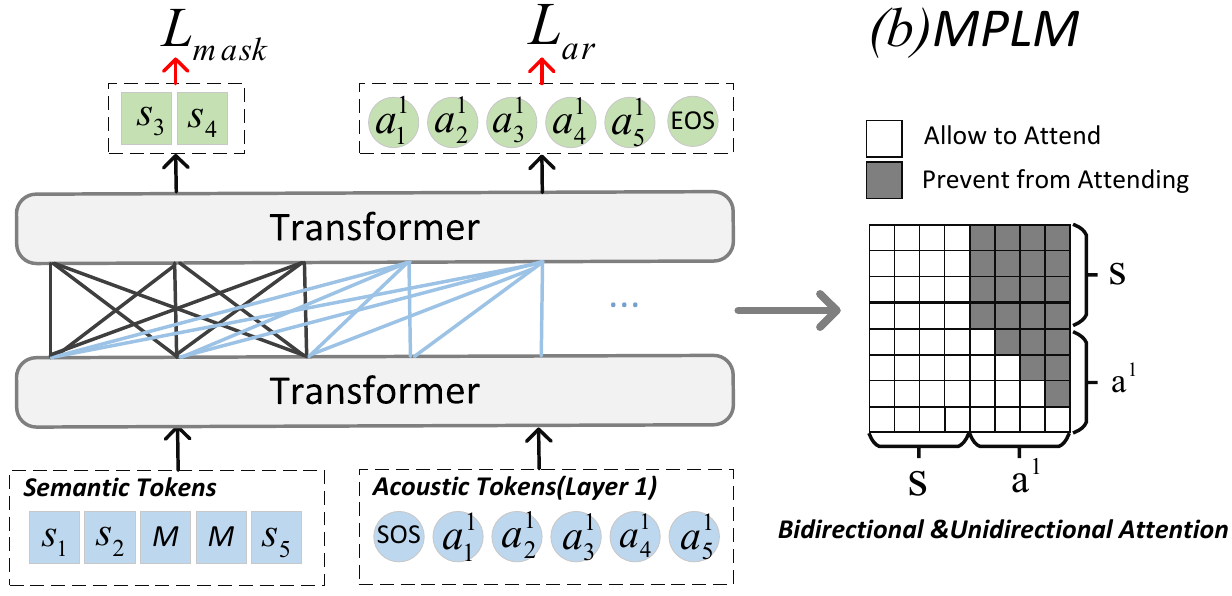}}
\end{minipage}
\textcolor[RGB]{180,180,180}{\rule{0.8\linewidth}{0.7pt}}
\begin{minipage}{0.8\linewidth}
    \subfigure{
      \includegraphics[width=1\columnwidth]{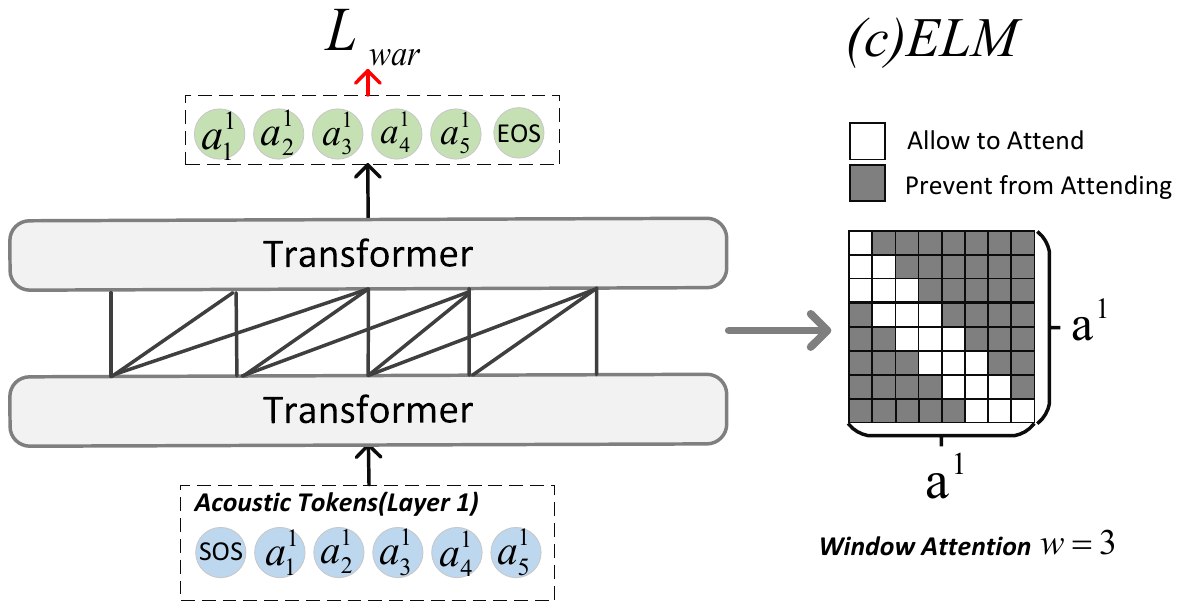}}
\end{minipage}
\vspace{-5pt}
\caption{The architecture of LM-VC. (a) The LM-VC model. (b) The masked prefix language model. (c) The external language model.}
\label{fig:lmvc}
\vspace{-15pt}
\end{figure}

\vspace{-10pt}
\section{Proposed Approach}
\vspace{-3pt}
\subsection{Overview}
\vspace{-5pt}

As shown in Fig.~\ref{fig:lmvc}(a), LM-VC incorporates three LMs: an MPLM, an ELM, and a PLM. 
Before language modeling, HuBERT~\cite{hsu2021hubert} and SoundStream~\cite{soundstream} are used to represent speech as semantic tokens $\mathbf{s} =\{s_1,s_2,...s_{T_s}\}$ and acoustic tokens $\mathbf{a} =\{a^1_1,a^2_1,...,a^L_1,a^1_2,...,a^L_{T_a}\}$, respectively. Here, $T_s$ and $T_a$ denote the sequence length, and $L$ represents the number of quantizers in SoundStream. 
Similar to AudioLM~\cite{audiolm}, LM-VC sequentially performs coarse and fine acoustic modeling. 

\textit{Coarse acoustic modeling}: The MPLM adopts the semantic tokens $\{\mathbf{s},\mathbf{\tilde{s}}\}$ from the source and target speaker speech, as well as the first-layer acoustic tokens $\tilde{\mathbf{a}}^1$ from target speaker speech. It autoregressively generates the acoustic tokens $\mathbf{a}^1$ of target speech, following the formulation $p(a^1_t|\tilde{\mathbf{s}},\mathbf{s},\tilde{\mathbf{a}}^1,\mathbf{a}^1_{1:t})$. In this process, the ELM performs $p(a^1_t|\mathbf{a}^1_{t-w:t})$ with window length $w$ and collaborates with the MPLM to generate speech. 

\textit{Fine acoustic modeling}: Taking the first-layer acoustic tokens as input, the PLM non-autoregressively generates fine acoustic tokens layer by layer. The semantic and acoustic tokens from the source speech and target speaker are also regarded as the prompt of the PLM. This process can be formulated as $p(\mathbf{a}^l|\tilde{\mathbf{s}},\mathbf{s},\tilde{\mathbf{a}},\mathbf{a}^{1:l-1},l)$ with $l \in [2, L]$. Following the NAR model in VALL-E~\cite{valle}, the PLM is achieved by a multi-layer Transformer~\cite{transformer} with bidirectional attention, leading to fast and high-quality speech reconstruction.

Finally, SoundStream reconstructs waveform from the predicted acoustic tokens. In the two-stage modeling, coarse acoustic modeling plays a crucial role in recovering linguistic content and speaker timbre, while fine acoustic modeling contributes to the acoustic fine details. In LM-VC, we put more effort on coarse acoustic modeling as keeping the source content and the target speaker timbre is a challenging task in zero-shot VC. The MPLM and ELM, designed for coarse acoustic modeling, are introduced in the following sections.

\vspace{-13pt}
\subsection{Masked Prefix Language Model}
\vspace{-3pt}
As just mentioned, obtaining high speaker similarity and preserving linguistic content are essential goals of the zero-shot VC. However, accomplishing these goals in LM is challenging since the linguistic content may get lost as the network deepens during multi-layer modeling, and the lengthy speech input makes learning contextual information harder, which causes unnatural pronunciation.
Furthermore, the semantic tokens still contain speaker-related information. And this inadequate decoupling causes the model to capture speaker timbre from both the target speaker speech and the source speech, thereby leading to low speaker similarity. Inspired by the advances in language modeling~\cite{GLM,ULM}, we introduce a masked prefix language model (MPLM) to address this issue.

As in Fig.~\ref{fig:lmvc} (b), MPLM is achieved by a multi-layer Transformer with two types of attention masks. To enhance the model's ability to learn contextual information and maintain the source content throughout the multi-layer modeling, MPLM employs a mask prediction strategy to restore masked tokens based on the surrounding context. Specifically, given a sequence of semantic tokens $\mathbf{s}=\{s_1,s_2,...s_{T_s}\}$, we randomly select several tokens as start indices at a ratio $r$, and spans of $l$ steps are masked by \textit{[M]} token. After masking, MPLM takes the corrupted semantic tokens $\mathbf{s}_{mask}$ as input and recovers the masked tokens. The right part of Fig.~\ref{fig:lmvc} (b) illustrates the self-attention mask used in MPLM. For the semantic tokens, a bidirectional attention mask allows them to attend to each other, enabling MPLM to capture contextual information from both directions. The negative log-likelihood loss, computed over masked tokens, can be defined as:
\vspace{-5pt}
\begin{equation} 
 \mathcal{L}_{mask} = -\log{\prod_{t\in M}p_{\mathrm{MPLM}}(s_t|\mathbf{s}_{mask},t)}.
\vspace{-5pt}
\end{equation}


For the acoustic generation, we employ the mask prediction strategy to make the model capture speaker timbre exclusively from the target speaker's speech, while extracting content from the corrupted semantic sequences. This strategy encourages the model to learn better contextual information and implicitly creates an information bottleneck in the semantic tokens to facilitate disentanglement. Moreover, during training, we do not explicitly use a speech clip as the acoustic prompt. Instead, MPLM leverages the previous acoustic sequence $\mathbf{a}^1_{1:t-1}$ as acoustic prompts to capture fine-grained speaker information and autoregressively generate $a^1_{t}$. In this process, we use unidirectional attention to achieve a left-to-right LM objective, where the acoustic token $a^1_t$ only attends to the previous sequence $\mathbf{a}^1_{1:t-1}$ and the semantic prefix $\mathbf{s}_{mask}$. The loss is
\vspace{-7pt}
\begin{equation} 
 \mathcal{L}_{ar} = -\log{\prod^{T_{a}-1}_{t = 0}p_{\mathrm{MPLM}}(a^1_t|\mathbf{a}^1_{1:t-1},\mathbf{s}_{mask},t)},
\vspace{-5pt}
\end{equation}
where $T_a$ represents the sequence length of acoustic tokens. During training, the semantic recovery and acoustic generation are performed simultaneously as $\mathcal{L}_{mask}+\mathcal{L}_{ar}$.

\vspace{-10pt}
\subsection{External Language Model}

In the generation process of MPLM, the generation diversity inherent in the sampling of the language model sometimes leads to unexpected results. This issue can be further amplified by autoregressive propagation, resulting in unnatural pronunciation and even speech quality degradation. 
Lack of guidance in the generation process, MPLM is hard to prevent this issue. Inspired by the phenomenon observed in contrastive predictive coding (CPC), previous studies~\cite{cpc,wav2vec2} have shown that adjacent speech frames within a speech segment of a specific length share the same local context, such as phoneme-related information. Such characteristic allows speech frames to be predicted by frames from previous time steps. As shown in Fig.\ref{fig:lmvc} (c), we introduce an external language model (ELM) to capture the local acoustic relations and provide contextual guidance during the generation process. With a similar architecture to the MPLM, the ELM employs window attention~\cite{liu2021swin} with shifted window to encode local contextual information and predict the distribution $p(a^1_t|\mathbf{a}^1_{t-w:t-1})$ with window length $w$. The objective of ELM can be defined as:
\vspace{-5pt}
\begin{equation} 
 \mathcal{L}_{war} = -\log{\prod^{T_{a}-1}_{t = 0}p_{\mathrm{ELM}}(a^1_t|\mathbf{a}^1_{t-w:t-1},t)}.
\vspace{-7pt}
\end{equation}

During training, we separately train the MPLM and ELM. In inference, the ELM collaborates with the MPLM to generate acoustic tokens conditioned on the local context of the preceding acoustic tokens. This collaboration is achieved through \textit{shallow fusion}~\cite{shallow1} with fusion weight $\lambda$:
\vspace{-4pt}
\begin{equation} 
\begin{split}
 a^1_t = argmax_{a^1_t}[\log{p_{\mathrm{MPLM}}(a^1_t|\mathbf{a}^1_{1:t-1},\mathbf{\tilde{a}^1},\mathbf{s},\mathbf{\tilde{s}},t)}\\+\lambda\log{p_{\mathrm{ELM}}(a^1_t|\mathbf{a}^1_{t-w:t-1},t)}].
\end{split}
\vspace{-13pt}
\end{equation}
Note that shallow fusion is wildly used in ASR~\cite{shallow2} to improve linguistic correctness during acoustic decoding.

\vspace{-10pt}
\section{Experiments}
\label{sec:exp}
\vspace{-5pt}
\subsection{Experimental Setup}
\subsubsection{Corpus}
A mixed dataset comprising 1,400 hours of LibriTTS~\cite{LibriTTS} and an internal dataset are used to train LM-VC and the SoundSteam codec~\cite{soundstream}. To extract semantic tokens, we incorporate an open-source HuBERT\footnote{https://github.com/bshall/hubert}, which is trained on LibriSpeech~\cite{librispeech}. For zero-shot testing, a set of 500 testing pairs is selected from VCTK~\cite{VCTK}, CMU Arctic~\cite{CMU-Arctic}, and EMIME~\cite{emime}, 
each with a source and target speaker utterance. 


\subsubsection{Implement details}
The SoundStream codec has 6 quantizer layers with a 1024 codebook size, representing a 24KHz waveform in 12.5ms frame length. The HuBERT compresses a 16KHz waveform into semantic tokens with 20ms frame length. For LM-VC, we employ the same decoder-only Transformers for MPLM, ELM, and PLM, with 12 layers, 16 attention heads, embedding dimension of 1024, feed-forward layer dimension of 4096, and dropout of 0.1, as in AudioLM~\cite{audiolm}. 
During training, the training length is capped at 10s. MPLM and PLM are trained using 8 A100 80G GPUs with a batch size of 12 per GPU for 600K steps, while ELM has a batch size of 20. We use the AdamW optimizer with a learning rate of $5 \times 10^{-4}$ for MPLM and ELM and $1 \times 10^{-4}$ for PLM. Exponential decay updates the learning rate after each epoch, using a decay ratio 0.986. In MPLM, mask ratio $r$ ranges from $0.02$ to $0.04$, and span $l$ is set to 10. Window length $w$ of the ELM is set to 20, whose temporal granularity is 250ms. And the fusion weight $\lambda$ is set to 0.3.

\subsubsection{Comparison systems}
Two representative VC systems are compared. We first implement a variant of AudioLM~\cite{audiolm} for VC (AuidoLM-VC), which uses semantic and first-layer acoustic tokens for coarse acoustic modeling. For a fair comparison, AudioLM-VC and LM-VC use the same PLM for fine acoustic modeling and both are trained on the same dataset. We also include a recent state-of-the-art VC system YourTTS~\cite{yourtts} with an open-source checkpoint as another comparison system.


\subsubsection{Evaluation metrics}
The mean opinion score (MOS) subjectively measures speech naturalness (NMOS) and speaker similarity (SMOS). We randomly select 120 testing pairs for subjective evaluations, involving a group of 15 listeners. For objective evaluations, a neural network-based system~\cite{nnmos} is used to measure speech quality (P-QMOS). Word error rate (WER) measured by an ASR model\footnote{https://github.com/wenet-e2e/wenet/tree/main/examples/librispeech/s0} indicates the speech intelligibility. Following previous work~\cite{FragmentvcAVLin2021}, speaker accuracy (ACC) is calculated by an SV model~\cite{ECAPA_TDNN} to determine if the converted speech matches the target speaker. Converted samples can be found in \href{https://kerwinchao.github.io/lmvc}{\url{https://kerwinchao.github.io/lmvc}}.

\vspace{-14pt}
\subsection{Experimental Results}
\vspace{-2pt}

\subsubsection{Subjective and objective results}

As presented in Table~\ref{exp:mos&obj}, compared with AudioLM-VC, the proposed model LM-VC achieves superior results for speech naturalness while getting higher P-QMOS and better WER. This indicates that the proposed model effectively preserves the linguistic content of source speech and produces natural speech. From the SMOS results, it can be found that LM-VC is effective in capturing the target speaker's timbre in the zero-shot VC task. Similar results are observed in terms of ACC. Additionally, LM-VC makes an obvious improvement compared to YourTTS regarding speech naturalness and speaker similarity. The P-QMOS and ACC also indicate the superiority of LM-VC.

Further assessment of LM-VC is conducted with ablations on the MPLM and ELM, as shown at the bottom of Table~\ref{exp:mos&obj}. Specifically, we replace the MPLM with the autoregressive prefix LM, as in AudioLM~\cite{audiolm}, forming the model \textit{w/o MPLM}. We observe a noticeable decrease in all evaluation metrics when the MPLM is discarded. This indicates that the MPLM, trained with mask prediction behavior, effectively enhances the zero-shot performance in capturing target speaker timbre while maintaining the source linguistic content. Furthermore, excluding the ELM from the inference process in the model \textit{w/o ELM} leads to a performance decrease in both NMOS and SMOS, highlighting the beneficial role of ELM in zero-shot VC. The objective metrics also report similar results.

\begin{table}[]

\setlength{\tabcolsep}{0.6mm}
\renewcommand\arraystretch{1.3}
\centering
\vspace{-5pt}
\caption{Results of subjective and objective evaluations. NMOS and SMOS are calculated with 95$\%$ confidence intervals}
\vspace{-5pt}
\label{exp:mos&obj}
\begin{tabular}{lccccc}
\hline
 Model &  NMOS $(\uparrow)$ & SMOS $(\uparrow)$  & P-QMOS $(\uparrow)$  & WER  $(\downarrow)$ & ACC $(\uparrow)$ \\ \hline
 GroundTruth & -  & -  & 4.56  & 1.64 &  1.000 \\ 
 YourTTS & 3.57$\pm$0.09  & 3.35$\pm$0.10  & 4.02 &  2.16 & 0.640   \\
 AudioLM-VC & 3.63$\pm$0.09  & 3.67$\pm$0.09  & 4.29 &  2.37 & 0.919   \\ \hline
 \makecell[l]{LM-VC} & \textbf{3.90$\pm$0.09}  & \textbf{3.82$\pm$0.10}  & \textbf{4.32} & \textbf{2.13}  & \textbf{0.951}  \\
 $\quad$ w/o MPLM&  3.83$\pm$0.10 & 3.69$\pm$0.09  & 4.31     & 2.31  & 0.926 \\
 $\quad$ w/o ELM&  3.86$\pm$0.09 & 3.73$\pm$0.08  & 4.32  & 2.20  & 0.942 \\ \hline
\end{tabular}
\vspace{-15pt}
\end{table}

\begin{figure}[htb]	
\vspace{-5pt}
	\centering
    	\includegraphics[width=0.85\linewidth]{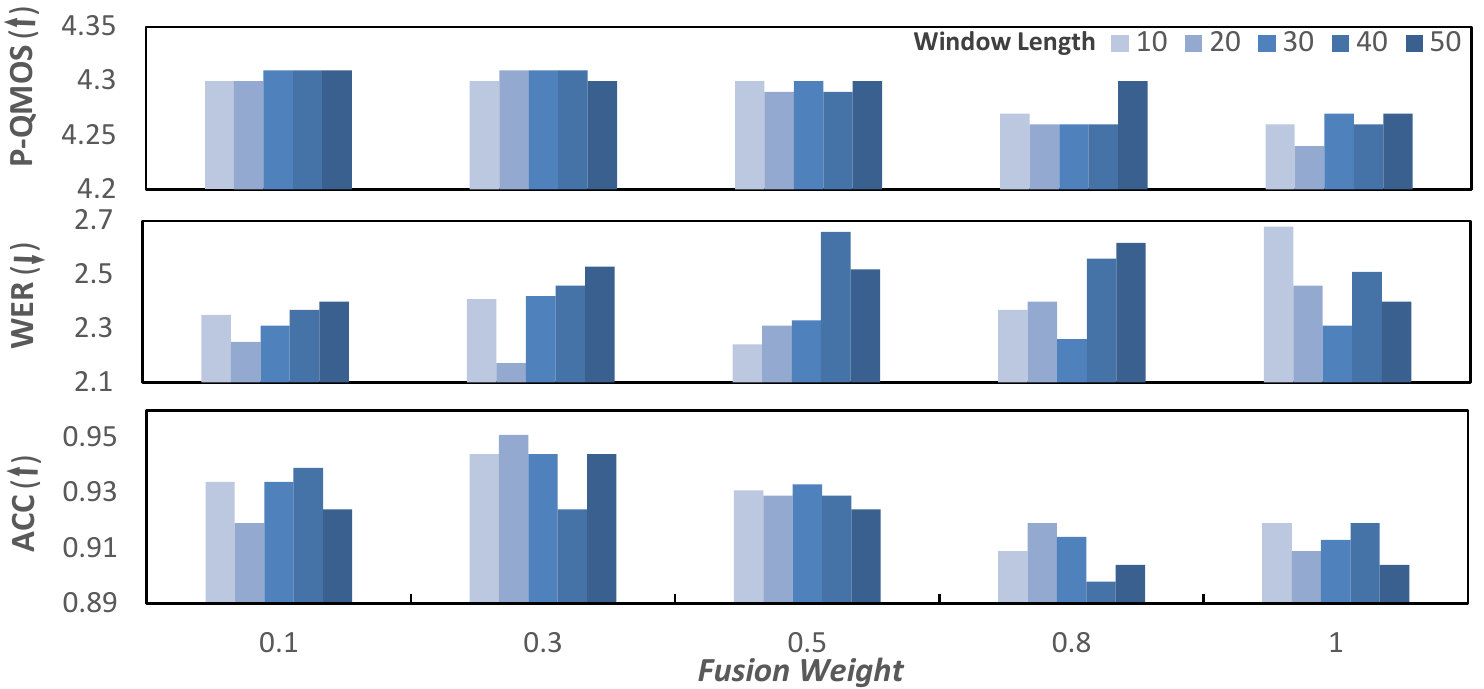}
 \vspace{-10pt}	
 \caption{Validation of ELM under different window lengths and fusion weights}
	\label{fig:windows}
 \vspace{-5pt}
\end{figure}

\begin{figure}[htb]	
	\centering
	\includegraphics[width=0.85\linewidth]{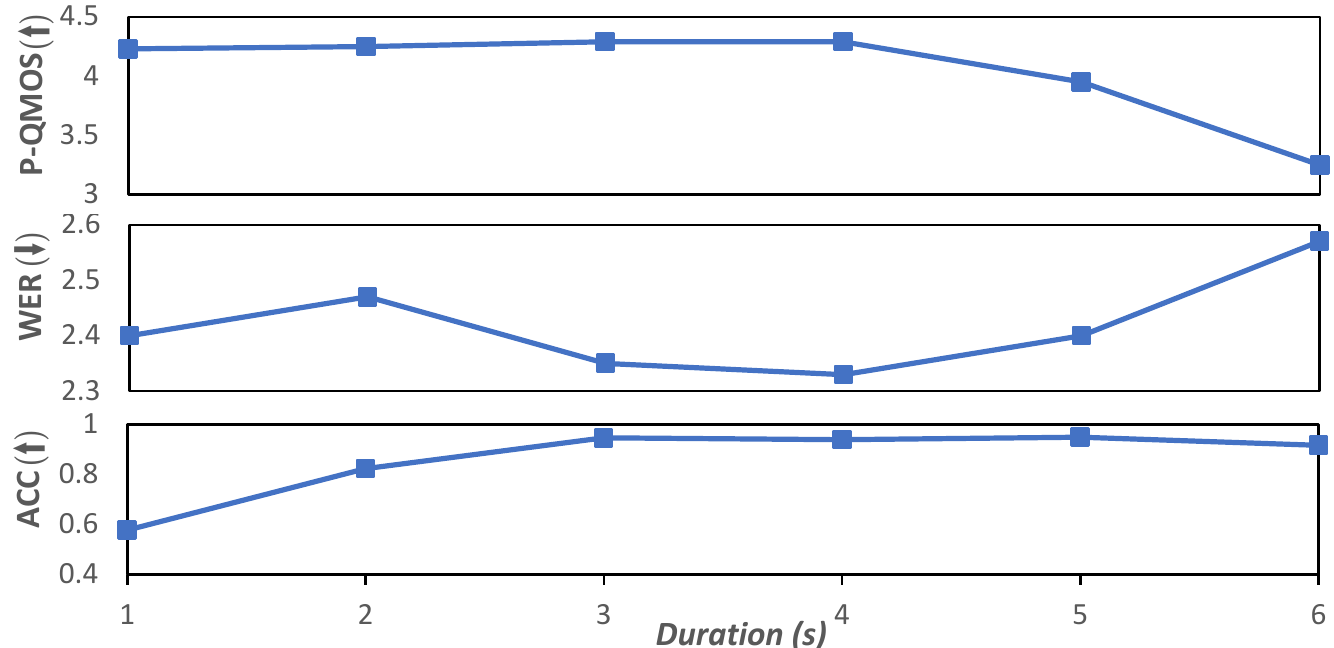}
 \vspace{-10pt}
	\caption{Zero-shot performance under the different duration of speaker prompt}
	\label{fig:duration}
 \vspace{-13pt}
\end{figure}

\subsubsection{Validation of ELM}
To examine the efficacy of ELM in LM-VC, we implement multiple ELM with varying window lengths ($w=10,20,30,40,50$) and fusion weights ($\lambda=0.1,0.3,0.5,0.8,1$). The objective results in Fig.~\ref{fig:windows} show that LM-VC initially improves across all three aspects as the window length increases, but subsequently begins to decline. When considering different fusion weights, we observe subtle variations in the trend. Additionally, LM-VC achieves optimal performance under distinct fusion weights. Notably, the setup of ($w=20, \lambda=0.3$), which covers
the common range of consonant-vowel syllables~\cite{duration}, exhibits the best performance. 



\subsubsection{Varying duration}
We further evaluate the zero-shot performance using different duration of target speaker utterances (1 $\sim$ 6s). As depicted in Fig.~\ref{fig:duration}, the results are generally affected by the duration of the speaker speech. For the extreme cases, e.g., 1 $\sim$ 2s, the model exhibits poor performance in WER and ACC due to insufficient prompt information. But from 3 to 4s, the objective scores noticeably improve in terms of speech intelligibility and speaker accuracy. Beyond 5s, the P-QMOS and WER get worse. This can be attributed to the long prompt (9 $\sim$ 12s) composed of the source and speaker speech. During speech generation of LM-VC, the input length can reach 1170$\sim$1569 tokens, which is close to or exceeds the maximum training length (10s, 1300 tokens). With bigger GPU memory or specific-designed structures~\cite{dao2023flashattention2,yu2023megabyte}, this problem may be alleviated with the increased training length. Meanwhile, leveraging longer prompt more effectively~\cite{Retrievrevise} is also a promising solution to break such length restriction.

\vspace{-5pt}
\section{CONCLUSIONS}

In this letter, we propose an LM-based zero-shot VC. Specifically, \textit{LM-VC} adopts a two-stage framework.
For coarse acoustic modeling, an MPLM is adopted in a mask prediction manner to enhance context learning and facilitate better disentanglement. Additionally, an ELM collaborates with MPLM to generate the target speech while ensuring speech generation stability. Finally, a non-autoregressive PLM reconstructs the fine acoustic tokens from the coarse acoustic tokens. Experiments demonstrate the superiority of LM-VC. The proposed LM-VC can easily \textbf{clone} desired speaker timbre with only 3 seconds of speech, simplifying any-to-any VC for applications like voice assistants, dubbing, and other speech generation scenarios. But it may cause potential risks in misuse, such as generating fake audio impersonating a specific speaker~\cite{yamagishi2021asvspoof}.

We have to point out that the current zero-shot VC approaches mainly consider the accurate delivery of the speaker timbre while lacking the modeling of other important speaker characteristics, such as accent and prosody, which are critical for identifying specific speakers. Besides, the out-of-domain problem still exists. Even trained with 60k hours of speech, LM-VC cannot ensure high speaker similarity for utterances with accents, strong emotions, or unseen recording environments, a similar limitation also pointed out in VALL-E~\cite{valle}.

\newpage

\bibliographystyle{IEEEtran}
\bibliography{ref.bib}

\begin{thebibliography}{10}
\providecommand{\url}[1]{#1}
\csname url@samestyle\endcsname
\providecommand{\newblock}{\relax}
\providecommand{\bibinfo}[2]{#2}
\providecommand{\BIBentrySTDinterwordspacing}{\spaceskip=0pt\relax}
\providecommand{\BIBentryALTinterwordstretchfactor}{4}
\providecommand{\BIBentryALTinterwordspacing}{\spaceskip=\fontdimen2\font plus
\BIBentryALTinterwordstretchfactor\fontdimen3\font minus
  \fontdimen4\font\relax}
\providecommand{\BIBforeignlanguage}[2]{{%
\expandafter\ifx\csname l@#1\endcsname\relax
\typeout{** WARNING: IEEEtran.bst: No hyphenation pattern has been}%
\typeout{** loaded for the language `#1'. Using the pattern for}%
\typeout{** the default language instead.}%
\else
\language=\csname l@#1\endcsname
\fi
#2}}
\providecommand{\BIBdecl}{\relax}
\BIBdecl

\bibitem{autovcqian2019autovc}
K.~Qian, Y.~Zhang, S.~Chang, X.~Yang, and M.~Hasegawa-Johnson, ``{AutoVC}:
  Zero-shot voice style transfer with only autoencoder loss,'' in
  \emph{International Conference on Machine Learning (ICML)}, 2019, pp.
  5210--5219.

\bibitem{speechsplit}
K.~Qian, Y.~Zhang, S.~Chang, M.~Hasegawa-Johnson, and D.~Cox, ``Unsupervised
  speech decomposition via triple information bottleneck,'' in
  \emph{International Conference on Machine Learning (ICML)}, 2020, pp.
  7836--7846.

\bibitem{mediumvc}
Y.~Gu, Z.~Zhang, X.~Yi, and X.~Zhao, ``{MediumVC}: Any-to-any voice conversion
  using synthetic specific-speaker speeches as intermedium features,''
  \emph{Arxiv}, 2021.

\bibitem{PPGSun2016PhoneticPF}
L.~Sun, K.~Li, H.~Wang, S.~Kang, and H.~Meng, ``{Phonetic Posteriorgrams for
  Many-to-One Voice Conversion without Parallel Data Training},'' in
  \emph{International Conference on Multimedia and Expo (ICME)}, 2016, pp.
  1--6.

\bibitem{nansy}
H.-S. Choi, J.~Lee, W.~Kim, J.~Lee, H.~Heo, and K.~Lee, ``Neural analysis and
  synthesis: Reconstructing speech from self-supervised representations,'' in
  \emph{Neural Information Processing Systems(NeurIPS)}, 2021, pp.
  16\,251--16\,265.

\bibitem{FragmentvcAVLin2021}
Y.~Y. Lin, C.~M. Chien, J.~hao Lin, H.~yi~Lee, and L.-S. Lee, ``{FragmentVC}:
  Any-to-any voice conversion by end-to-end extracting and fusing fine-grained
  voice fragments with attention,'' in \emph{International Conference on
  Acoustics, Speech and Signal Processing (ICASSP)}, 2021, pp. 5939--5943.

\bibitem{retriever}
D.~Yin, X.~Ren, C.~Luo, Y.~Wang, Z.~Xiong, and W.~Zeng, ``Retriever: Learning
  content-style representation as a token-level bipartite graph,'' in
  \emph{International Conference on Learning Representations (ICLR)}, 2021.

\bibitem{speaking}
G.~Maimon and Y.~Adi, ``Speaking style conversion with discrete self-supervised
  units,'' \emph{Arxiv}, 2022.

\bibitem{INchou2019oneshot}
J.~chieh Chou and H.-Y. Lee, ``One-shot voice conversion by separating speaker
  and content representations with instance normalization,'' in
  \emph{International Speech Communication Association (Interspeech)}, 2019,
  pp. 664--668.

\bibitem{wang2023multilevel}
Z.~Wang, L.~Xue, Q.~Kong, L.~Xie, Y.~Chen, Q.~Tian, and Y.~Wang, ``Multi-level
  temporal-channel speaker retrieval for robust zero-shot voice conversion,''
  \emph{Arxiv}, 2023.

\bibitem{VQMIVC}
D.~Wang, L.~Deng, Y.~T. Yeung, X.~Chen, X.~Liu, and H.~Meng, ``{VQMIVC}: Vector
  quantization and mutual information-based unsupervised speech representation
  disentanglement for one-shot voice conversion,'' in \emph{International
  Speech Communication Association (Interspeech)}, 2021, pp. 1344--1348.

\bibitem{MAP}
J.~Wang, J.~Li, X.~Zhao, Z.~Wu, S.~Kang, and H.~Meng, ``Adversarially learning
  disentangled speech representations for robust multi-factor voice
  conversion,'' \emph{Arxiv}, 2021.

\bibitem{avqvc}
H.~Tang, X.~Zhang, J.~Wang, N.~Cheng, and J.~Xiao, ``{AVQVC}: One-shot voice
  conversion by vector quantization with applying contrastive learning,'' in
  \emph{Conference on Acoustics, Speech and Signal Processing (ICASSP)}, 2022,
  pp. 4613--4617.

\bibitem{contrastive}
J.~Ebbers, M.~Kuhlmann, T.~Cord-Landwehr, and R.~Haeb-Umbach, ``Contrastive
  predictive coding supported factorized variational autoencoder for
  unsupervised learning of disentangled speech representations,'' in
  \emph{International Conference on Acoustics, Speech and Signal Processing
  (ICASSP)}, 2021, pp. 3860--3864.

\bibitem{audiolm}
Z.~Borsos, R.~Marinier, D.~Vincent, E.~Kharitonov, O.~Pietquin, M.~Sharifi,
  O.~Teboul, D.~Grangier, M.~Tagliasacchi, and N.~Zeghidour, ``{AudioLM}: a
  language modeling approach to audio generation,'' \emph{Arxiv}, 2022.

\bibitem{musiclm}
A.~Agostinelli, T.~I. Denk, Z.~Borsos, J.~Engel, M.~Verzetti, A.~Caillon,
  Q.~Huang, A.~Jansen, A.~Roberts, M.~Tagliasacchi, M.~Sharifi, N.~Zeghidour,
  and C.~Frank, ``{MusicLM}: Generating music from text,'' \emph{Arxiv}, 2023.

\bibitem{valle}
C.~Wang, S.~Chen, Y.~Wu, Z.~Zhang, L.~Zhou, S.~Liu, Z.~Chen, Y.~Liu, H.~Wang,
  J.~Li, L.~He, S.~Zhao, and F.~Wei, ``Neural codec language models are
  zero-shot text to speech synthesizers,'' \emph{Arxiv}, 2023.

\bibitem{spearTTS}
E.~Kharitonov, D.~Vincent, Z.~Borsos, R.~Marinier, S.~Girgin, O.~Pietquin,
  M.~Sharifi, M.~Tagliasacchi, and N.~Zeghidour, ``{Speak, Read and Prompt}:
  High-fidelity text-to-speech with minimal supervision,'' \emph{ArXiv}, 2023.

\bibitem{hsu2021hubert}
W.-N. Hsu, B.~Bolte, Y.-H.~H. Tsai, K.~Lakhotia, R.~Salakhutdinov, and
  A.~Mohamed, ``{HuBERT}: Self-supervised speech representation learning by
  masked prediction of hidden units,'' \emph{Transactions on Audio, Speech, and
  Language Processing}, vol.~29, pp. 3451--3460, 2021.

\bibitem{liu2021swin}
Z.~Liu, Y.~Lin, Y.~Cao, H.~Hu, Y.~Wei, Z.~Zhang, S.~Lin, and B.~Guo, ``{Swin
  Transformer}: Hierarchical vision transformer using shifted windows,'' in
  \emph{International conference on computer vision (ICCV)}, 2021, pp.
  10\,012--10\,022.

\bibitem{shallow1}
C.~Gulcehre, O.~Firat, K.~Xu, K.~Cho, L.~Barrault, H.-C. Lin, F.~Bougares,
  H.~Schwenk, and Y.~Bengio, ``On using monolingual corpora in neural machine
  translation,'' \emph{Arxiv}, 2015.

\bibitem{yourtts}
E.~Casanova, J.~Weber, C.~D. Shulby, A.~C. Junior, E.~G{\"o}lge, and M.~A.
  Ponti, ``{YourTTS}: Towards zero-shot multi-speaker {TTS} and zero-shot voice
  conversion for everyone,'' in \emph{International Conference on Machine
  Learning (ICML)}, 2022, pp. 2709--2720.

\bibitem{soundstream}
N.~Zeghidour, A.~Luebs, A.~Omran, J.~Skoglund, and M.~Tagliasacchi,
  ``{SoundStream}: An end-to-end neural audio codec,'' \emph{Transactions on
  Audio, Speech, and Language Processing}, vol.~30, pp. 495--507, 2021.

\bibitem{transformer}
A.~Vaswani, N.~Shazeer, N.~Parmar, J.~Uszkoreit, L.~Jones, A.~N. Gomez, L.~u.
  Kaiser, and I.~Polosukhin, ``Attention is all you need,'' in \emph{Neural
  Information Processing Systems (NerurIPS)}, vol.~30, 2017.

\bibitem{GLM}
Z.~Du, Y.~Qian, X.~Liu, M.~Ding, J.~Qiu, Z.~Yang, and J.~Tang, ``{GLM}: General
  language model pretraining with autoregressive blank infilling,'' in
  \emph{Association for Computational Linguistics (ACL)}, 2022, pp. 320--335.

\bibitem{ULM}
L.~Dong, N.~Yang, W.~Wang, F.~Wei, X.~Liu, Y.~Wang, J.~Gao, M.~Zhou, and H.-W.
  Hon, ``Unified language model pre-training for natural language understanding
  and generation,'' in \emph{Neural Information Processing Systems (NerurIPS)},
  vol.~32, 2019.

\bibitem{cpc}
A.~v.~d. Oord, Y.~Li, and O.~Vinyals, ``Representation learning with
  contrastive predictive coding,'' \emph{Arxiv}, 2018.

\bibitem{wav2vec2}
A.~Baevski, Y.~Zhou, A.~Mohamed, and M.~Auli, ``wav2vec 2.0: A framework for
  self-supervised learning of speech representations,'' in \emph{Neural
  Information Processing Systems (NerurIPS)}, vol.~33, 2020, pp.
  12\,449--12\,460.

\bibitem{shallow2}
R.~Cabrera, X.~Liu, M.~Ghodsi, Z.~Matteson, E.~Weinstein, and A.~Kannan,
  ``Language model fusion for streaming end to end speech recognition,''
  \emph{Arxiv}, 2021.

\bibitem{LibriTTS}
H.~Zen, V.~Dang, R.~Clark, Y.~Zhang, R.~J. Weiss, Y.~Jia, Z.~Chen, and Y.~Wu,
  ``{LibriTTS}: A corpus derived from {LibriSpeech} for text-to-speech,'' in
  \emph{International Speech Communication Association (Interspeech)}, 2019,
  pp. 1526--1530.

\bibitem{librispeech}
V.~Panayotov, G.~Chen, D.~Povey, and S.~Khudanpur, ``{LibriSpeech}: an {ASR}
  corpus based on public domain audio books,'' in \emph{International
  conference on acoustics, speech and signal processing (ICASSP)}, 2015, pp.
  5206--5210.

\bibitem{VCTK}
C.~Veaux, J.~Yamagishi, and K.~MacDonald, ``{CSTR VCTK} corpus: English
  multi-speaker corpus for {CSTR} voice cloning toolkit.''\hskip 1em plus 0.5em
  minus 0.4em\relax University of Edinburgh. The Centre for Speech Technology
  Research (CSTR), 2016.

\bibitem{CMU-Arctic}
J.~Kominek and A.~W. Black, ``The {CMU Arctic} speech databases,'' in \emph{5th
  ISCA Workshop on Speech Synthesis (SSW 5)}, 2004, pp. 223--224.

\bibitem{emime}
M.~Wester, ``The {EMIME} bilingual database,'' The University of Edinburgh,
  Tech. Rep., 2010.

\bibitem{nnmos}
X.~Shu, Y.~Chen, C.~Shang, Y.~Zhao, C.~Zhao, Y.~Zhu, C.~Huang, and Y.~Wang,
  ``Non-intrusive speech quality assessment with a multi-task learning based
  subband adaptive attention temporal convolutional neural network,'' in
  \emph{International Speech Communication Association (Interspeech)}, 2022,
  pp. 3298--3302.

\bibitem{ECAPA_TDNN}
B.~Desplanques, J.~Thienpondt, and K.~Demuynck, ``{ECAPA-TDNN}: Emphasized
  channel attention, propagation and aggregation in {TDNN} based speaker
  verification,'' in \emph{International Speech Communication Association
  (Interspeech)}, 2020, pp. 3830--3834.

\bibitem{duration}
M.~Steinschneider, K.~V. Nourski, and Y.~I. Fishman, ``Representation of speech
  in human auditory cortex: is it special?'' \emph{Hearing research}, vol. 305,
  pp. 57--73, 2013.

\bibitem{dao2023flashattention2}
T.~Dao, ``Flashattention-2: Faster attention with better parallelism and work
  partitioning,'' \emph{Arxiv}, 2023.

\bibitem{yu2023megabyte}
L.~Yu, D.~Simig, C.~Flaherty, A.~Aghajanyan, L.~Zettlemoyer, and M.~Lewis,
  ``Megabyte: Predicting million-byte sequences with multiscale transformers,''
  \emph{Arxiv}, 2023.

\bibitem{Retrievrevise}
R.~Das, M.~Zaheer, D.~Thai, A.~Godbole, E.~Perez, J.~Y. Lee, L.~Tan,
  L.~Polymenakos, and A.~McCallum, ``Case-based reasoning for natural language
  queries over knowledge bases,'' in \emph{Association for Computational
  Linguistics (ACL)}, 2021, pp. 9594--9611.

\bibitem{yamagishi2021asvspoof}
J.~Yamagishi, X.~Wang, M.~Todisco, M.~Sahidullah, J.~Patino, A.~Nautsch,
  X.~Liu, K.~A. Lee, T.~Kinnunen, N.~Evans \emph{et~al.}, ``Asvspoof 2021:
  accelerating progress in spoofed and deepfake speech detection,''
  \emph{Arxiv}, 2021.

\end{thebibliography}

\end{document}